\documentclass[10pt,article]{IEEEtran}
\usepackage[utf8]{inputenc}
\usepackage{graphicx}
\usepackage{pifont}
\usepackage{caption}
\usepackage{subcaption}
\usepackage{listings}
\usepackage{hyperref}

\lstdefinestyle{mystyle}{
    basicstyle=\ttfamily\footnotesize,
    breakatwhitespace=false,         
    breaklines=true,                 
    captionpos=b,                    
    keepspaces=true,                 
    numbers=left,                    
    numbersep=5pt,                  
    showspaces=false,                
    showstringspaces=false,
    showtabs=false,                  
    tabsize=2,
    linewidth=.99\linewidth,
    xleftmargin=0.3cm, % bump it over a hair to stay in linewidth
    frame=tbrl, %t: top, r, b, l 
}
\usepackage[htt]{hyphenat} % hyphenation in texttt

\title{Interconnect Bandwidth Heterogeneity on AMD MI250x and Infinity Fabric
\thanks{
Sandia National Laboratories is a multimission laboratory managed and operated by National Technology and Engineering Solutions of Sandia, LLC., a wholly owned subsidiary of Honeywell International, Inc., for the U.S. Department of Energy's National Nuclear Security Administration under contract DE-NA-0003525.
This research was supported by the Exascale Computing Project (17-SC-20-SC), a collaborative effort of the U.S. Department of Energy Office of Science and the National Nuclear Security Administration.
This research used resources of the Oak Ridge Leadership Computing Facility at the Oak Ridge National Laboratory, which is supported by the Office of Science of the U.S. Department of Energy under Contract No. DE-AC05-00OR22725.
The author would like to thank James Elliot and Simon Garcia de Gonzalo of Sandia National Labs for their feedback and guidance.
}
}
\author{Carl Pearson\\
cwpears@sandia.gov\\
Sandia National Labs\\
Albquerque, NM 87185}

\author{\IEEEauthorblockN{Carl Pearson} \\
\IEEEauthorblockA{\textit{Sandia National Laboratories}\\
Albuquerque, NM, USA \\
cwpears@sandia.gov}
}

\begin{document}

\maketitle

\begin{abstract}
Demand for low-latency and high-bandwidth data transfer between GPUs has driven the development of multi-GPU nodes.
Physical constraints on the manufacture and integration of such systems has yielded heterogeneous intra-node interconnects, where not all devices are connected equally.
The next generation of supercomputing platforms are expected to feature AMD CPUs and GPUs.
This work characterizes the extent to which interconnect heterogeneity is visible through GPU programming APIs on a system with four AMD MI250x GPUs, and provides several insights for users of such systems.
\end{abstract}

\iffalse
This paper presents the first published evaluation of communication primitives on AMD's CDNA 2 architecture GPUs and Infinity Fabric 3 interconnect. Through a series of benchmarks, the performance of the HIP API on Frontier's heterogeneous multi-GPU system is evaluated. Results show that the interconnect heterogeneity manifests at the HIP API level as significant bandwidth differences between HIP devices depending on which devices are participating. Furthermore, NUMA effects were not observed between CPU and GPU, and execution resources can be used to increase communication bandwidth when GPUs are idle. The results of this work provide insights into communication performance and resource utilization on heterogeneous multi-GPU systems.
\fi

\section{Introduction}

Multi-GPU nodes are a common feature of high-performance computing systems due to their density and power-efficiency.
For  suitable workloads, GPUs offer more floating-point operations per watt, and more operations per second available within a single node.
Physical constraints place limits on the number and bandwidth of chip-to-chip interconnects.
Therefore, multi-CPU and multi-GPU nodes may be organized with heterogeneous interconnects between different pairs of components to maximize the amount of compute available in the node.
Systems with AMD x86 CPUs and AMD GPUs are expected to be forerunners in the upcoming generation of supercomputing systems.
Frontier~\cite{olcf2022frontier} is an example of such a system; a familiar design, with a single CPU connected to four GPUs by AMD's proprietary Infinity Fabric interconnect.

This work contributes 
\begin{itemize}
\item the first published relationship between communication primitives and achieved bandwidth on AMD's CDNA 2 GPUs and Infinity Fabric 3 interconnect.
\end{itemize}
Furthermore, it makes the following observations from the results:
\begin{itemize}
\item like previous multi-GPU systems, interconnect heterogeneity manifests at the HIP API level as significant bandwidth differences between HIP devices depending on which devices are participating,
\item unlike previous multi-CPU, multi-GPU systems, NUMA effects were not observed between CPU and GPU,
\item if GPUs are idle during communication, execution resources can be used to achieve higher bandwidth.
\end{itemize}

The rest of this paper is organized as follows.
Section~\ref{sec:methodology} discusses the measurement methods,
Sec.~\ref{sec:results} discusses the results,
Sec~\ref{sec:related} discusses related work,
and Sec.~\ref{sec:conclusion} concludes.

\section{Methodology}
\label{sec:methodology}

\subsection{Evaluation System}

AMD provides the ROCm software platform~\cite{amd2022rocm} for GPU-accelerated systems.
It includes C++ language extensions, kernel language, and runtime APIs (together known as HIP) necessary to interact with AMD GPUs, as well as higher-level libraries.
This work uses the HIP runtime API and kernel language to evaluate point-to-point CPU/GPU communication bandwidth on a flagship HPC node:  Crusher~\cite{olcf2022crusher} at the Oak Ridge Leadership Computing Facility (OLCF), which has nodes identical to Frontier.

\begin{table}[ht]
\centering
\begin{tabular}{c | c} 
\textbf{Feature}   & \textbf{Description}                \\
\hline
CPU                    & AMD EPYC 7A53                       \\
GPU                    & 4x AMD MI250x (2x GCD)             \\
GCD                    & 4x2 AMD CDNA2               \\
CPU-GCD                & Infinity Fabric 72+72 GB/s        \\
Intra-GPU (``quad'')   & Infinity Fabric 200+200 GB/s      \\
Inter-GPU (``dual'')   & Infinity Fabric 100+100 GB/s      \\
Inter-GPU (``single'') & Infinity Fabric 50+50 GB/s          \\
ROCm                   & 5.4.0                               \\
GPU Driver             & 5.16.9.22.20.7582                   \\
OS                     & SUSE Linux Enterprise Server 15 SP3 \\
Kernel                 & Linux 5.3.18                        \\
\end{tabular}
\caption{
OLCF Crusher node summary.
Bandwidths are given as the sum of each direction.
Figure~\ref{fig:crusher-node} summarizes the node topology.
``Quad'', ``dual'', and ``single'' refer to the number of connections drawn in that figure.
}
\label{tab:crusher-node}
\end{table}

\begin{figure}[ht]
\centering
\includegraphics[width=\linewidth]{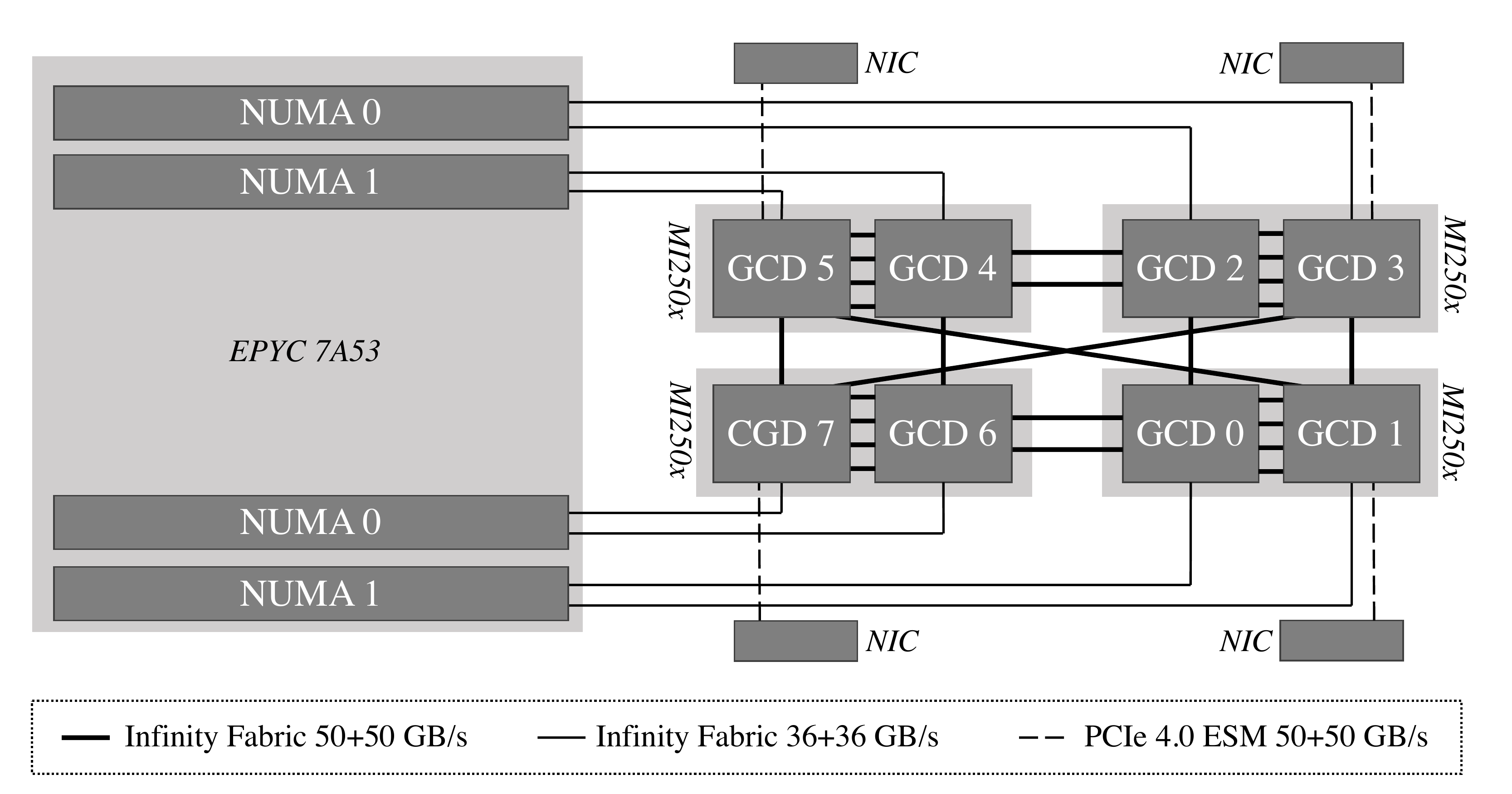}
\caption{
Crusher node block-diagram, adapted from~\cite{olcf2022crusher}.
Each of the four AMD MI250x GPUs has two graphics compute dies (GCDs), each of which is an addressable HIP device acting as an individually-programmable GPU.
The system features PCIe 4.0, as well as bidirectional 50 + 50 GB/s and 36 + 36 GB/s Infinity Fabric interconnects.
}
\label{fig:crusher-node}
\end{figure}

Table~\ref{tab:crusher-node} and Figure~\ref{fig:crusher-node} summarize the topology of the node.
Each node has a single AMD EPYC 7A53 CPU with four AMD MI250x GPUs (the ``host'' and ``devices'' respectively).
Each MI250x has two graphics compute dies (GCDs), each of which can be addressed as a unique HIP device.
CPUs and GCDs are connected by different kinds of Infinity Fabric links~\cite{amd2022cdna2}.

The in-package infinity fabric offers 200+200 GB/s bidirectional bandwidth to neighboring GCD (``quad'')
There are also 8 lanes of inter-package Infinity Fabric, for 400+400 GB/s total.
On the evaluation platform, this is allocated in two 100+100 GB/s (``dual'') and one 50+50 GB/s (``single'') connection to other GCDs and one 36+36 GB/s coherent connection to an L3 slice on the CPU.
There is also a 50+50 GB/s PCIe 4.0 Extended Speed Mode connection to NIC not investigated in this work.

\vspace{-2mm}
\subsection{Allocation Types}

Table~\ref{tab:combinations} summarizes the buffer types and transfer methods in this work.
All allocations are coarse-grained, i.e. they do not support any coherency during GPU kernel execution.

Device allocations are produced with \texttt{hipMalloc}.
They may be used for for an explicit transfer, or mapped into another HIP device using \texttt{hipPeerAccessEnable} so a kernel executing on that device can access the data implicitly.

Host pinned allocations are produced by \texttt{hipHostMalloc} with \texttt{hipHipHostMallocNumaUser} and \texttt{hipHipHostMallocNonCoherent} flags.
The buffer can be used in an explicit transfer, or accessed implicitly by a GPU kernel with \texttt{hipHostGetDevicePointer}.
The data in this allocation may not be paged by the host.

Host pageable allocations are produced with the system allocator (e.g., \texttt{malloc}).
This buffer can be used in an explicit transfer; internally HIP runtime will stage the data through a pinned buffer so the GPU DMA engine can read it from physical memory.

Managed allocations are produced with \texttt{hipMallocManaged} and \texttt{hipMemAdviseSetCoarseGrain}.
This buffer may be transparently accessed by the GPU and the CPU with various degrees coherency.

\vspace{-2mm}
\subsection{Transfer Methods}
\label{sec:transfer_methods}

\textit{Explicit} transfers use \texttt{hipMemcpyAsync}.
A buffer is created on the source and destination, and \texttt{hipMemcpyAsync} is invoked to move data from source to destination.

\textit{Implicit} transfers use load and store operations in a GPU kernel to move data from source to destination.
For \textit{implicit mapped} transfers between host and device, a buffer is created on the host, and the device writes to or reads from that buffer.
For device-device transfers, the buffer is created on the destination device, and the source writes to it.
For \textit{implicit managed} transfers, a single managed allocation is produced.
It is prefetched to the source device, and then modified from the destination device.
\texttt{HSA\_XNACK=1} is set in the environment, causing pages to be migrated from the source to destination device.
The \texttt{cpu\_write} function uses an OpenMP-parallel loop to write 64-bit elements to the buffer with sequential values equal to their index plus a fixed constant.
The \texttt{gpu\_write} kernel does the same using a large HIP grid, where threads make coalesced accesses to the buffer.
The \texttt{gpu\_read} kernel instead reads those values.

\textit{Prefetch} transfers use \texttt{hipMemPrefetchAsync} on a managed buffer so the data is resident on a particular device.
The data will then be local to that device if that device makes future accesses.

\begin{table}[ht]
\centering
\begin{tabular}{c c c c} 
\textbf{Host Buffer} & \textbf{Device Buffer} & \textbf{Transfer} & \textbf{Direction} \\
\hline
malloc        & hipMalloc              & explicit & H2D, D2H      \\
hipHostMalloc & hipMalloc              & explicit & H2D, D2H      \\
hipHostMalloc & --                     & implicit & H2D, D2H      \\
 --           & hipMalloc              & explicit & D2D           \\
 --           & hipMalloc              & implicit & D2D           \\
\multicolumn{2}{c}{hipMallocManaged}  & implicit & H2D, D2H, D2D \\
\multicolumn{2}{c}{hipMallocManaged}  & prefetch & H2D, D2H, D2D \\
\end{tabular}
\caption{
Combinations of buffer types and transfer methods evaluated in this work.
``Explicit'' transfers use \texttt{hipMemcpyAsync}.
``Implicit'' transfers use load/store operations on a buffer pointer in a GPU kernel.
``D2H,'' ``H2D','' and ``D2D'' mean device-to-host, host-to-device, and device-to-device respectively.
}
\label{tab:combinations}
\end{table}

\vspace{-3mm}
\subsection{Measurement}

Measurements are implemented in the Comm{\textbar{}}Scope package at \url{github.com/c3sr/comm_scope}.
The Google Benchmark support library~\cite{google2022benchmark} is used to drive the various benchmarks.
It chooses the number of measurements iterations such that the operation in question executes for at least one second, at least once, and fewer than one billion times.
For these benchmarks, the fastest (GPU-to-GPU implicit writes) were invoked $\approx$59000 times, and the slowest (prefetches) twice.

In the setup phase, NUMA affinity for host allocations is enforced as necessary.
\texttt{hipDeviceReset} is used to reset the devices to discard any state that may have accumulated from past benchmark runs.
\texttt{hipSetDevice} controls the active device(s) while necessary buffers are created and filled to ensure a physical memory mapping.
%\texttt{hipEvent\_t}s track the time of asynchronous operations.

During the benchmark iterations, the benchmark state is reset by a combination of device affinities, cache flushes, memory fills, and prefetches to get the buffers to a known state.
For asynchronous operations, the start HIP event is recorded, the operation is invoked in the default stream, and the stop event is recorded.
For synchronous operations, an \texttt{std::chrono::high\_precision} clock is recorded, the operation is invoked, and the clock is recorded again to measure the wall time.
This process is repeated until the Google Benchmark harness' conditions are satisfied.
During teardown, all resources and NUMA bindings are released.

\section{Results}
\label{sec:results}

\begin{figure*}[htbp]
     \centering
     \hfill
     \begin{subfigure}[t]{0.34\textwidth}
         \centering
         \includegraphics[width=\textwidth]{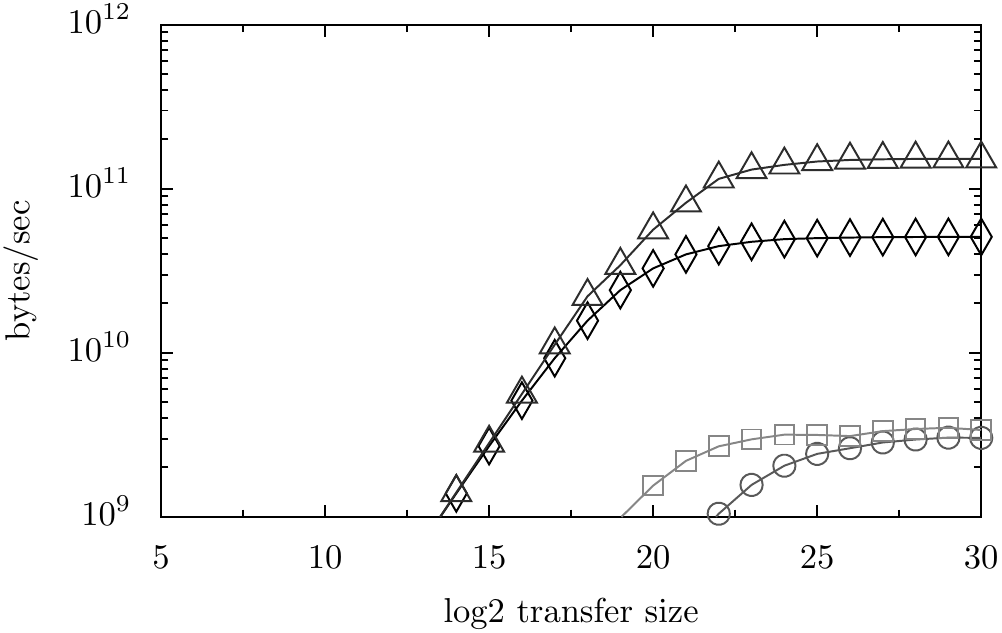}
\caption{
GCD-GCD bandwidth across ``quad'' Infinity Fabric links, e.g. GCD 0 to GCD 1.
}
         \label{fig:crusher_gpu-to-gpu_quad}
     \end{subfigure}
     \begin{subfigure}[t]{0.318\textwidth}
         \centering
         \includegraphics[width=\textwidth]{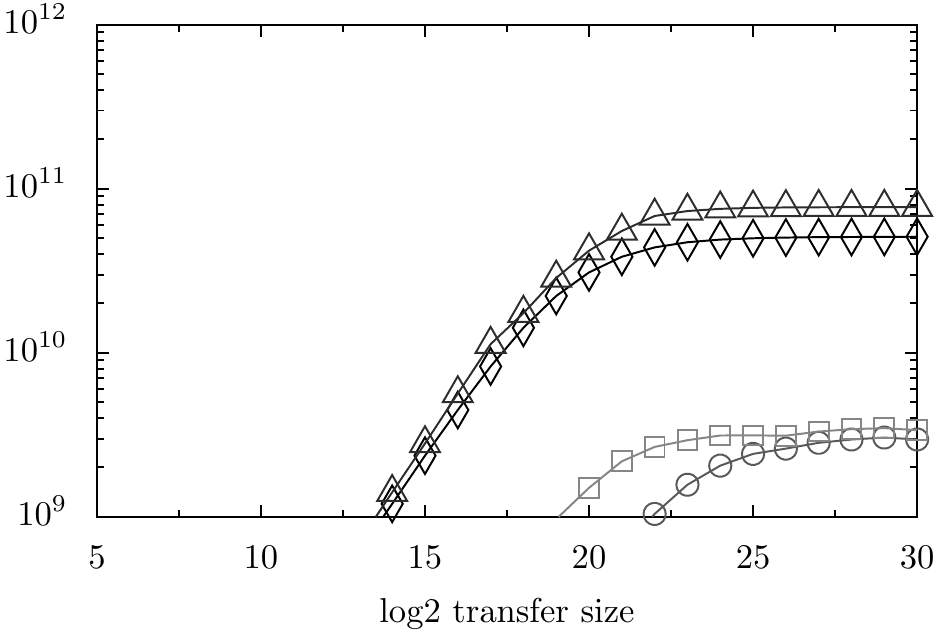}
\caption{
GCD-GCD bandwidth across ``dual'' Infinity Fabric links, e.g. GCD 0 to GCD 6.
}
         \label{fig:crusher_gpu-to-gpu_dual}
     \end{subfigure}
     \hfill
     \begin{subfigure}[t]{0.318\textwidth}
         \centering
         \includegraphics[width=\textwidth]{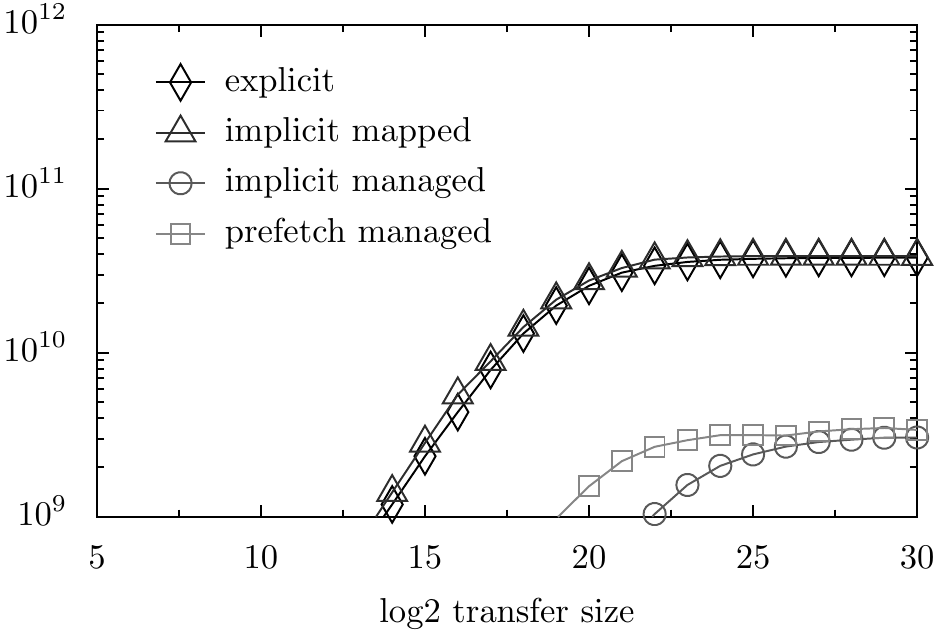}
\caption{
GCD-GCD bandwidth across ``single'' Infinity Fabric links, e.g. GCD 0 to GCD 2.
}
         \label{fig:crusher_gpu-to-gpu_single}
     \end{subfigure}
\caption{
Unidirectional GCD-to-GCD bandwidth measured on the OLCF Crusher system.
}
\label{fig:crusher_gpu-to-gpu}
\end{figure*}

\begin{figure*}[htbp]
\centering
\begin{subfigure}[b]{0.508\textwidth}
\centering
\includegraphics[width=\textwidth]{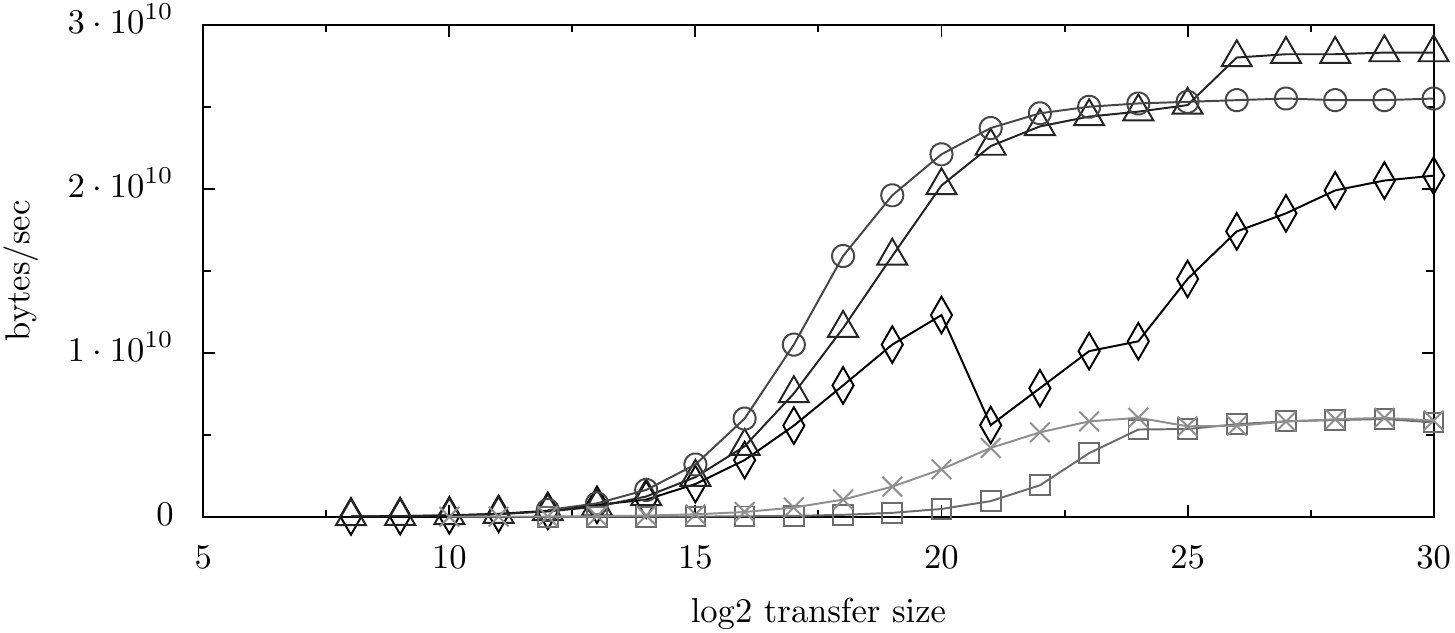}
\caption{
Host-to-device bandwidth.
}
\label{fig:crusher_host-to-gpu}
\end{subfigure}
\hfill
\begin{subfigure}[b]{0.482\textwidth}
\centering
\includegraphics[width=\textwidth]{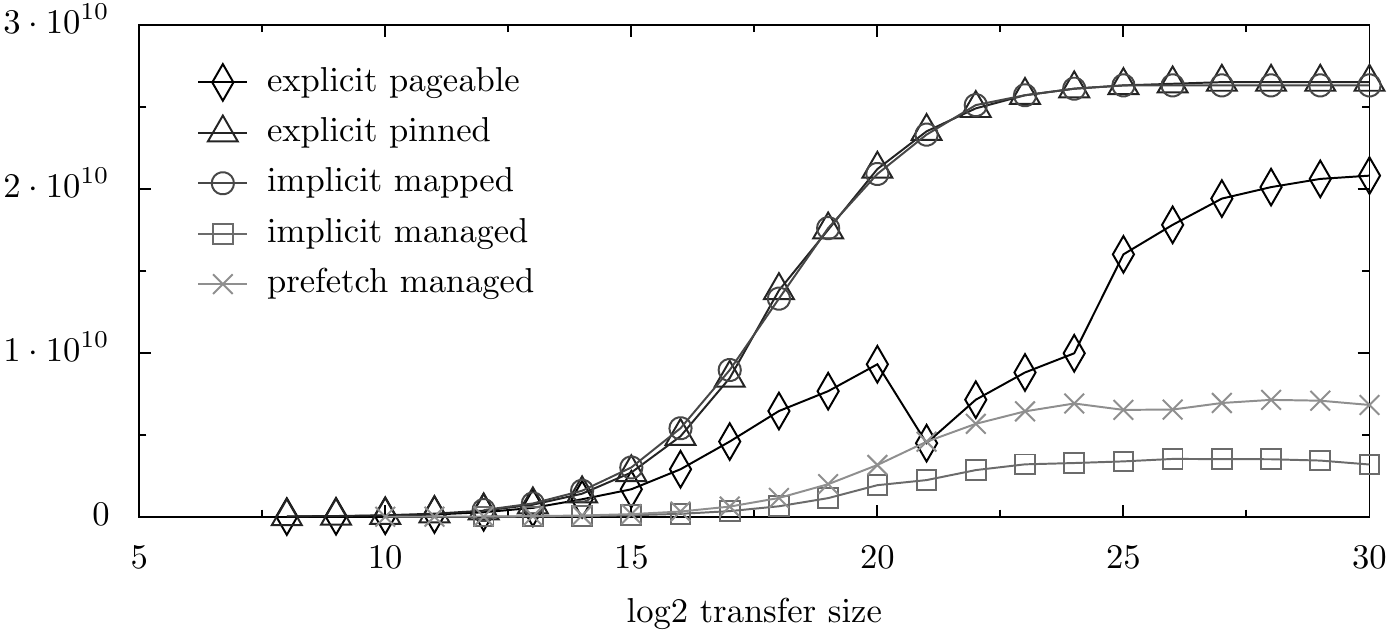}
\caption{
Device-to-host bandwidth.
}
\label{fig:crusher_gpu-to-host}
     \end{subfigure}
\caption{
Unidirectional NUMA-node / GPU bandwidth on OLCF Crusher node.
All pairs of NUMA-node and HIP device have the same performance characteristics.
Results for NUMA 0 and GCD 0 shown.
}
\label{fig:crusher_host-gpu}
\end{figure*}

Figures~\ref{fig:crusher_gpu-to-gpu_quad},~\ref{fig:crusher_gpu-to-gpu_dual},~and~\ref{fig:crusher_gpu-to-gpu_single} show the measured GCD-to-GCD bandwith for GCDs 0-1 (quad-link, intra-GPU), 0-6 (dual link), and 0-2 (single-link) respectively.
Figure~\ref{fig:crusher_host-to-gpu}~and~\ref{fig:crusher_gpu-to-host} show the measured CPU-to-GCD and GCD-to-CPU bandwidth, respectively.
The primary observed effect is that the transfer method has an enormous impact on achieved bandwidth over fixed hardware.
Only implicit access by GPU kernels are able to saturate every interconnect in the system.

\vspace{-3mm}
\subsection{Managed Memory Prefetch Bandwidth}

The \texttt{hipMemPrefetchAsync} hint can be used to prefetch managed data to a specific device in anticipation of future access there.
It is several orders of magnitude (up to $1630\times$, $47\times$ for 1 GiB transfers) slower than the fastest transfer method.
It is too slow to elucidate any interconnect heterogeneity, and will be excluded from discussion in the sections below.

\vspace{-3mm}
\subsection{Transfer Method Effects}

The effect of the transfer method is most significant for the fastest interconnects.
Table~\ref{tab:peak-d2d} shows what fraction of the theoretical peak bandwidth is achieved for each transfer method for different GCD-GCD connections.
For faster intra-GPU interconnects, achieved bandwidth varies by $3\times$, while for the slower single-linked GCDs all transfer methods are roughly equivalent, with dual-link GCDs somewhere inbetween.
Implicit transfers between mapped buffers are always able to saturate the link.
This suggests that using the GPU's computing resources to speed up transfers may be more effective than using slower DMA methods, especially if the GPU is idle during the transfer.

\begin{table}[ht]
\centering
\begin{tabular}{c | c c c } 
\textbf{Transfer} & \multicolumn{3}{c}{\textbf{Interconnect Type}}\\
         & ``quad'' & ``dual'' & ``single'' \\
\hline                                    
explicit         & 0.25  & 0.51  & 0.76  \\
implicit mapped  & 0.77  & 0.77  & 0.78  \\
implicit managed & 0.74  & 0.76  & 0.76  \\
prefetch managed & 0.016 & 0.032 & 0.064 \\
\hline
\textit{Peak GB/s} & \textit{200} & \textit{100} & \textit{50} \\
\end{tabular}
\caption{
Fraction of theoretical peak bandwidth for 1 GiB device/device transfers
}
\label{tab:peak-d2d}
\end{table}

While the 50 GB/s GCD-GCD link is slow enough to make all transfer methods equal, the even slower CPU-GCD link is still fast enough to reveal the slowing effect of staging pagable transfers through pinned memory.
Figures~\ref{fig:crusher_gpu-to-gpu}~and~\ref{fig:crusher_host-gpu} show that such transfers are $5\times$ slower than pinned allocations in the worst case.

\subsection{Device/Device Topology Effects}

Device-to-device topology effects are especially apparent for the fastest transfer methods.
Implicit access to a mapped allocation on a different device is able to utilize $\approx$75\% of the available bandwidth, yielding 153 GB/s within a GPU, 77 GB/s between dual-link GPUs, and 39 GB/s between single-link GPUs.
Explicit transfers seem to have a ceiling of 51 GB/s, so while they are able to saturate the single-link GPU to approximately the same bandwidth as implicit access (38 GB/s), they are only able to generate 51 GB/s across the much higher-bandwidth intra- and inter-GPU connections.
This suggests the DMA engine in CDNA2 may only be able to generate 51 GB/s of memory traffic for a given transfer.

\subsection{Host/Device Topology Effects}

Despite different NUMA regions being linked with different GCDs, there are no observable NUMA effects on transfer bandwidth (i.e., bandwidth is the same regardless of NUMA region or GCD involved).
Thus, Figs.~\ref{fig:crusher_host-to-gpu}~and~\ref{fig:crusher_gpu-to-host} are representative of any host-device pairing.
While these results suggest that NUMA affinity of host allocations does not matter, it may become more relevant if multiple transfers are in flight simultaneously.

These experiments cannot reveal whether this homogeneity is a result of the CPU memory subsystem or the GPU interconnects, since the fastest CPU/GPU transfers are slower than the slowest 38 GB/s GPU-GPU transfers, so any GPU-GPU interconnect effects would not be visible.

\vspace{-2mm}
\subsection{Anisotropic Effects}

CPU-to-GCD (Fig.~\ref{fig:crusher_host-to-gpu}, ``implicit managed'') transfers are much faster than GCD-to-CPU (\ref{fig:crusher_gpu-to-host}, ``implicit managed'').
For the former, the GPU is initiating writes to CPU-resident pages and vis-versa for the latter.
This is the only substantial anisotropic transfer observed.

\vspace{-2mm}
\subsection{Related Work}
\label{sec:related}

Substantial prior work exists for investigating multi-GPU communication performance.
All prior work operates on widespread Nvidia platforms.
This work distinguishes itself by doing similar evaluations on AMD platforms.

Spafford, Meredith, and Vetter~\cite{spafford2011quantifying} identify NUMA-related bandwidth effects for CPU/GPU transfers.
This work makes similar measurements on a more modern platform and finds that it has no such NUMA effects.
Tallent et al.~\cite{tallent2018evaluating} investigate the performance of GPU-GPU data transfers.
While they focus on certain collective operations relevant to deep neural networks, they also meaure point-to-point transfer time at specific sizes.
Li et al.~\cite{li2019evaluating} present microbenchmarks of latency, bandwidth, and collective communications for PCIe 3.0, and Nvidia's NVLink, NVSwitch, and GPUDirect technologies.
Pearson et al.~\cite{pearson2019evaluating} do a similar evaluation for Nvidia NVLink interconnects and IBM POWER9 (excluding collective communications).
This work does a similar analysis of AMD's Infinity Fabric interconnect with AMD GPUs.
Alan and Ge~\cite{allen2021depth} and Chien, Peng, and Markidis~\cite{chien2019performance} present a characterization of the unified virtual memory system on Nvidia GPUs across microbenchmarks, synthetic kernels, and HPC workloads.
This work microbenchmarks achievable bandwidth for AMD's version of this technology.

\subsection{Conclusion and Future Work}
\label{sec:conclusion}

This paper presents GPU-GPU communication bandwidth measurements for a flagship AMD high-performance compute node.
The capabilities of the intra- and inter-GPU interconnects have outstripped the ability of the system software and DMA engines to saturate them.
If the workload requires GPU-to-GPU communication, it may be beneficial to implement it as implicit access between mapped buffers rather than relying on the DMA engine.
There were no observable NUMA effects impacting CPU/GPU transfers, which may simplify this dimension of multi-GPU application tuning.
The presented methodology is implemented in the Comm{\textbar{}}Scope package at \url{github.com/c3sr/comm_scope}.

Future systems like El Capitan that contain AMD processors are expected to feature integrated CPUs and GPUs, with substantially different performance characteristics despite the same ROCm programming development platform. 
These systems may feature even higher-bandwidth interconnects and tighter integration, which further emphasize distinctions between transfer methods and impact the performance of implicit accesses between different devices.

The evaluation in this paper is limited to unidirectional point-to-point transfers between CPUs and GPUs using HIP APIs.
There are significant additional transfer types of interest to the community, including simultaneous (including bi-directional and collective), GPU-NIC, and off-node.

\bibliographystyle{IEEEtran}
\bibliography{main}

\end{document}